\documentclass[aps,prb,twocolumn,reprint,showpacs]{revtex4-2}
\usepackage{graphicx}
\usepackage{epstopdf}
\usepackage{dcolumn}
\usepackage{bm}
\usepackage[utf8]{inputenc}
\usepackage[T1]{fontenc}
\usepackage{mathptmx}
\usepackage{mathpazo}
\usepackage[colorlinks=true,urlcolor=blue,linkcolor=blue,citecolor=blue]{hyperref}
\usepackage{amsmath}
\usepackage{amssymb}
\usepackage{eurosym}
\usepackage{txfonts}
\usepackage{enumerate}
\usepackage{enumitem}
\usepackage{gensymb}
\usepackage{ulem}
\usepackage{cancel}
\usepackage{xcolor}
\usepackage[greek,english]{babel}

\begin{document}
\title{Anisotropy of linear magnetoresistance in Kagome metal ZrV$_6$Sn$_6$}

    \author{Yifan Deng,$^{\textcolor{blue}1,\textcolor{blue}2}$}
    \author{Ming Cheng,$^{\textcolor{blue}1,\textcolor{blue}2}$}
	\author{Lanxin Liu,$^{\textcolor{blue}1,\textcolor{blue}2}$}
    \author{Nan Zhou,$^{\textcolor{blue}1}$}	
    \author{Yu Zhao,$^{\textcolor{blue}1,\textcolor{blue}2}$}
    \author{Ruihuan Lan,$^{\textcolor{blue}1,\textcolor{blue}2}$}
    \author{Yongqiang Pan,$^{\textcolor{blue}1}$}
	\author{Wenhai Song,$^{\textcolor{blue}1}$}
    \author{Yuyan Han,$^{\textcolor{blue}3}$}
    \author{Xiaoguang Zhu,$^{\textcolor{blue}1}$}
    \author{Xuan Luo,$^{\textcolor{blue}1}$}
	\email[Correspondence authors: ]{xluo@issp.ac.cn}
    \author{Yuping Sun$^{\textcolor{blue}3,\textcolor{blue}1,\textcolor{blue}4}$}
	\email{ypsun@issp.ac.cn}

	\affiliation{
		$^{1}$Key Laboratory of Materials Physics, Institute of Solid State Physics, HFIPS, Chinese Academy of Sciences, Hefei 230031, China\\
		$^{2}$Science Island Branch of Graduate School, University of Science and Technology of China, Hefei 230026, China\\
		$^{3}$Anhui Key Laboratory of Low-Energy Quantum Materials and Devices, High Magnetic Field Laboratory, HFIPS, Chinese Academy of Sciences, Hefei, Anhui 230031, China\\
		$^{4}$Collaborative Innovation Center of Advanced Microstructures, Nanjing University, Nanjing 210093, China}

\date{\today}

\begin{abstract}

The Kagome lattice has attracted extensive attention due to the diverse magnetic properties and non-trivial electronic states generated by its unique atomic arrangement, which provides an excellent system for exploring macroscopic quantum behavior. Here, we report the anomalous transport properties in 166-type Kagome metal ZrV$_6$Sn$_6$ single crystals. The quadratic and linear magnetoresistance (LMR) can be observed depending on the directions of the field and the current. Integrating Hall resistivity and quantum oscillation measurements, we found that the LMR could match well with the Abrikosov model. However, this model encounters difficulties in explaining the anisotropy of the magnetoresistance. To solve the issue, we extrapolate the Abrikosov model to the case of two-dimensional linear dispersion. It was found that when the field is parallel to the linear dependence momentum, the quantized energy is $\epsilon_n^{\pm}$ = $\pm v\sqrt{p^2+2eHn/c}$, resulting in LMR. By contrast, when it is parallel to the non-linear dependence momentum, the energy is $\epsilon_n^{\pm}$ = $\pm v\sqrt{2eHn/c}$, without yielding LMR. Through the combination of experiment and theory, the modified Abrikosov model could interpret the macroscopic quantum transport in ZrV$_6$Sn$_6$ crystal. The present research provides a new perspective for understanding the LMR behavior.
\end{abstract}

\maketitle

Macroscopic quantum behavior refers to the manifestation of quantum-mechanical characteristics at the macroscopic scale, such as superconductivity \cite{SC1911,BCS,RMPSC}, superfluidity \cite{He1,He2,RMPBEC}, quantum tunneling \cite{QT1,QT2}, Shubnikov–de Haas (SdH) oscillations \cite{OSC1,OSC2,ZrV6Sn6} and quantum linear magnetoresistance (LMR) \cite{QLMR3d1,QLMR3d2,QLMR2d1,QLMR2d2}. They hold great significance in developing condensed matter physics theories and driving breakthroughs in innovative technologies. In the case of SdH oscillations and quantum LMR, the energy of electrons is redistributed under an external magnetic field, forming new discrete energy levels, namely Landau levels. According to the Abrikosov theory \cite{QLMR}, quantum LMR could be generated when electrons are filled in the lowest Landau level ($n$ = 0), which provides important insights for the study of the electron behavior at the quantum limit. The quantum LMR discussed by Abrikosov is in the case of three-dimensional (3D) linear dispersion, while the quantum LMR in two-dimensional (2D) linear dispersion deserves further research and discussion.

	Recently, the Kagome lattice has attracted considerable attention due to its special atomic arrangement, which is composed of regular hexagons and equilateral triangles. On one hand, antiferromagnetic Kagome insulators possess strong geometric frustration, and are regarded as candidates for realizing quantum spin liquids with fractionalized elementary excitations \cite{QSL2010,QSL2012,QSL2017}. On the other hand, Kagome metals exhibit non-trivial topological electronic band structures under the tight-binding approximation, such as Dirac dispersion, van Hove singularities (vHs), and flat bands \cite{kagomeband}. Until now, practical Kagome systems have been widely researched to find quantum spin liquid candidate \cite{ZnCu}, negative magnetism \cite{Co3Sn2S2} and orbital Hall effect \cite{Fe3Sn2}. The 135-type Kagome metal AV$_3$Sb$_5$ with Van Hove filling near the Fermi level \cite{135-1,135-2} has attracted much attention due to its unique superconductivity ground state \cite{135SC1,135SC2,135SC3,135multi} and unconventional density waves \cite{135CDW,135PDW}. The 166-type Kagome metal AT$_6$X$_6$ system exhibits diverse physical properties \cite{TbMn6Sn6,CsCr6Sb6,ScV6Sn6} due to its greater chemical tunability than the 135-type system. Therefore, the 166-type system has become a crucial lattice model for elucidating the mechanisms of macroscopic quantum behaviors. In recent research on (Ti,Zr,Hf)V$_6$Sn$_6$ \cite{ZrV6Sn6} about SdH oscillations, the nontrivial topological bands have been detected, which could be used to reveal more characteristics of quantum LMR. Compared to other 166-type kagome metals \cite{166Re1,166Re2,166Re3,166Re4,166Re5}, the ZrV$_6$Sn$_6$ exhibits no magnetism, enabling a more intrinsic investigation of the transport properties associated with its topological electronic structure. Furthermore, it hosts vHs and nodal lines near the Fermi level \cite{ZrV6Sn6}, making it an important platform for studying the topological band structures of 166-type systems.

\begin{figure*}
\includegraphics[width=0.8\linewidth]{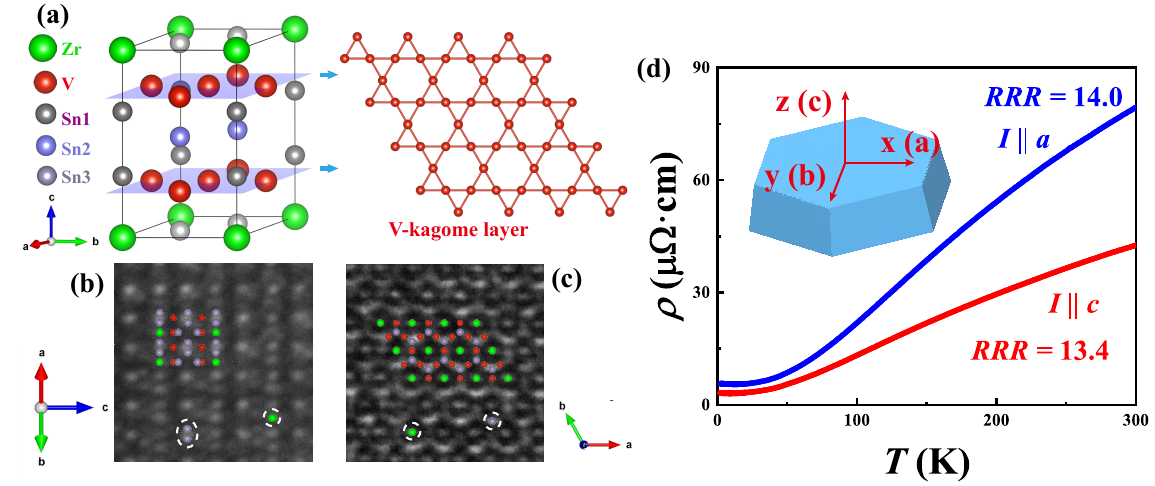}
\begin{center}
\caption{\label{1} (a) Crystal structure of ZrV$_6$Sn$_6$. (b), (c): HRTEM images with corresponding crystal structure of ZrV$_6$Sn$_6$ single-crystal. (d) Temperature-dependent resistivity \textit{$\rho$} of ZrV$_6$Sn$_6$ single-crystal when the current is parallel to the $a$-axis and $c$-axis, respectively. Inset: Sample measurement sketch with the $x$, $y$, and $z$ axes correspond to the $a$, $b$, and $c$ axes, respectively. }
\end{center}
\end{figure*}

In this paper, we report transport properties study on 166-type Kagome metal ZrV$_6$Sn$_6$. It was found that the quadratic and linear magnetoresistance have a close relationship with the direction of the field and current in ZrV$_6$Sn$_6$ crystals. In order to fully understand their intrinsic origin, we carefully investigated their Hall effects and SdH oscillations. By comparing and analyzing the magnetoresistance, Hall effect and SdH oscillation results, we consider that the LMR can be explained by the Abrikosov model \cite{QLMR}. We further developed Abrikosov's theory and demonstrated the results in a 2D linear dispersion system to explain the corresponding anisotropy of LMR in ZrV$_6$Sn$_6$.

ZrV$_6$Sn$_6$ single crystal samples were synthesized by Sn self-flux method. The detailed methods and characterizations of single crystal structure(X-ray diffraction (XRD), transmission electron microscopy (TEM) and Inductively Coupled Plasma (ICP)) are shown in Supplemental Material (SM) \cite{SM} (see also references \cite{kondo1,kondo2,Cp,hall1,hall2,166band1,166band2,166band3,166band4,LK,berry1,berry2,Bc} therein). The ZrV$_6$Sn$_6$ exhibits HfFe$_6$Ge$_6$-type structure with hexagonal space group P6/mmm (no. 191) as shown in Fig.\textcolor{blue}{1(a)}. Two V-layers (form Kagome layers), two Sn1-layers, Sn2-layer and Zr-Sn3-layer stack along the $c$-axis coordinate. High contrast regions of the TEM images in Fig.\textcolor{blue}{1(b)} and Fig.\textcolor{blue}{1(c)} can correspond to atomic positions within the ZrV$_6$Sn$_6$ unit cell. The temperature-dependence of resistivity for ZrV$_6$Sn$_6$ single crystal measured at zero field along $a$-axis and $c$-axis is demonstrated in Fig.\textcolor{blue}{1(d)}. The inset shows the corresponding relationship between the macroscopic crystal orientation, the coordinate axes of experimental measurement, and the directions of the crystal axes. The descending resistivity without any kinks indicates the metal electrical transport behavior without any transitions. The resistivity along the $a$-axis is almost twice that along the $c$-axis, with the residual resistivity ratio (RRR) of
    $\left[\rho\left( 300 K \right)-\rho\left( 2 K \right)\right]/\rho\left( 2 K \right) \approx$ 14.0 and 13.4, respectively.

To explore the anisotropic transport properties of ZrV$_6$Sn$_6$, we measured the field-dependence of the out-of-plane magnetoresistance ($MR\%$ = [$\rho$($\mu_0$$H$)-$\rho$(0)] / $\rho$(0) $\times$ 100\%) under different current and field configurations, which are shown in Fig.\textcolor{blue}{2(a)} and Fig.\textcolor{blue}{2(b)}. The insets of Fig.\textcolor{blue}{2(c)} and Fig.\textcolor{blue}{2(d)} represent Configuration A and Configuration B when $\theta_1$($\theta_2$) is equal to 0, respectively. Both configuration of magnetoresistance are positive values and increase with decreasing temperature. But the magnetoresistance along the $z$-direction is larger, about twice that along the $x$-direction. The result along the $x$-direction shows SdH oscillations at low temperatures. We will discuss it in the following text. And the result along the $z$-direction exhibits linear dependence rather than normal quadratic dependence , and it has also been found in other Kagome systems \cite{kagomeLMR1,kagomeLMR2,kagomeLMR3}. This effect gradually appears in high fields as the temperature decreases and extends to a lower field. The range of the LMR region at different temperatures will be discussed in Supplemental Material (SM) \cite{SM}.

In order to further explore the differences between these two configurations, we measured the magnetoresistance in different field directions while always keeping them perpendicular to the current. They are shown in Fig.\textcolor{blue}{2(c)} and Fig.\textcolor{blue}{2(d)}. As the field rotates, the $MR_{xx}$ gradually changes from quadratic to linear and then gradually returns to quadratic. The $MR_{zz}$ hardly changes regardless of how the field rotates. We fit the magnetoresistance by power-law ($MR$ = $a$$T^n$), and plot the angle-dependence of power $n$ in Fig.\textcolor{blue}{2(e)}. For $MR_{zz}$, $n$ always remains close to 1. For $MR_{xx}$, $n$ is 2 when the field is parallel to the $c$-axis, and it gradually approaches 1 when the field is parallel to the $b$-axis. From this, we draw the conclusion that the LMR occurs only when the field is parallel to the $ab$-plane. Several mechanisms for explaining the LMR have been proposed. We summarized these mechanisms and compared them with the results of our experiments.

\begin{figure*}
\includegraphics[width=1\linewidth]{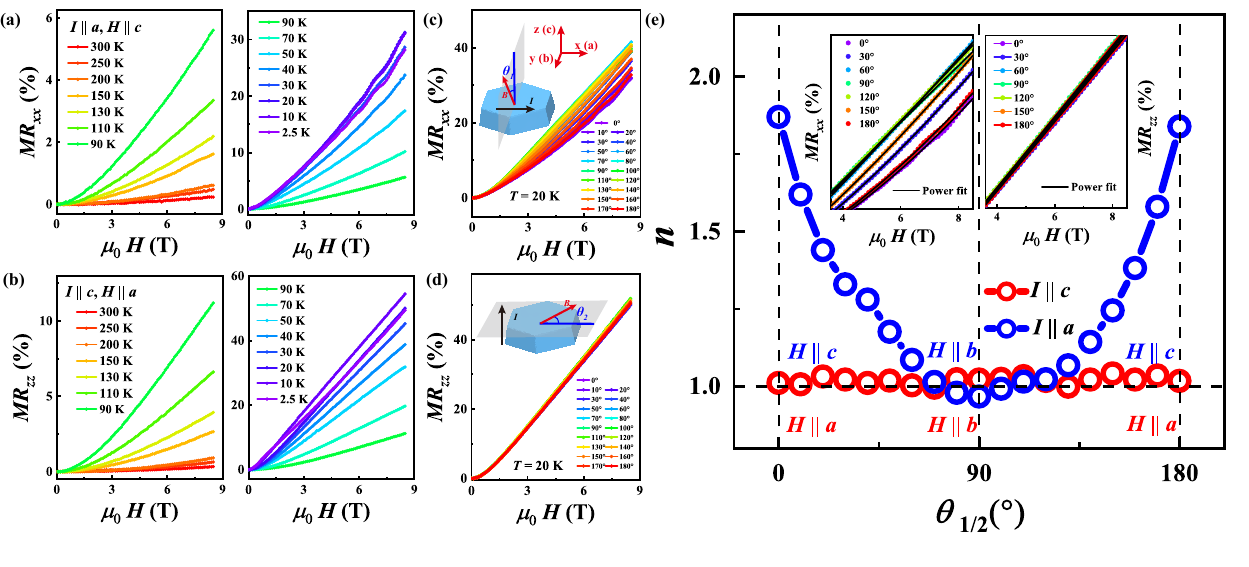}
\begin{center}
\caption{\label{1} The ($MR$) for ZrV$_6$Sn$_6$ single-crystal. (a), (b): The field-dependence of $MR$ at various temperatures. Configuration A (a): Current along $a$-axis and field along $c$-axis. Configuration B (b): Current along $c$-axis and field along $a$-axis. Inset: The temperature-dependence of crossover field ($B_c$). $B_c$ is the intersection point of the linear fitted by the high-field curve and the low-field curve in the field-dependence of derivative of $MR_{zz}$. (c), (d): The field-dependence of $MR$ at 20 K for the current along $a$-axis and $c$-axis , respectively, and field rotating within the normal plane of the current. Inset: Sample measurement sketch for different current and field configurations. (e): The angle-dependence of power $n$, which could fitted by $MR$ = $a$ $(T)^n$ in inset of (e). }
\end{center}
\end{figure*}

    Generally, the field-dependence of magnetoresistance will show a result of increasing quadratically at low field and reaching saturation at high field, and the power changes gradually from 2 to 0 (For the multi-band model, the system exhibits a similar regularity if the density of one type of carrier is much higher than that of the others)\cite{talkMR,TiB2}, and LMR could be observed in the intermediate region. The magnetoresistance can be expressed as \cite{halfMR1,halfMR2}:

\begin{equation}
		\frac{\Delta\rho}{\rho} \propto
	\begin{cases}
	(\mu H)^2, & \mu H < 1 \\
	C, & \mu H > 1
	\end{cases}
    \end{equation}

    Where $\mu$ is the carrier mobility. In the quasi-classical picture, the carriers rotate along the intersection line in the k-space of the plane perpendicular to B and the equipotential surface. When $\omega \tau<1$ and $H<1/\mu$ ($\tau$ is the carrier lifetime, $\omega$ is the cyclotron frequency of this field, and $\omega \tau=\mu H$), the carrier is scattered before completing one full circle, and the magnetoresistance will show a quadratic dependence. When $\omega \tau>1$ and $H>1/\mu$, the carrier can complete full circle, and the magnetoresistance will saturate as the field increases. Then, we measured the high field magnetoresistance up to 14 T of ZrV$_6$Sn$_6$, as shown in Fig.\textcolor{blue}{3(a)} for Configuration A and Fig.\textcolor{blue}{3(b)} for Configuration B. The oscillations in Configuration A become more prominent, and it is discussed thoroughly in SM \cite{SM}. The magnetoresistance of Configuration B extends to a larger value and keeps linear. There are still no signs of saturation in magnetoresistance even after reaching $H^*$($H^*=1/\mu$) in Configuration B, and $\mu$ can be obtained from Hall measurements in Fig.\textcolor{blue}{S4} \cite{SM}. Furthermore, as shown in Fig.\textcolor{blue}{3(c)}, the Kohler rule \cite{Kohler} at different temperatures for Configuration B does not fall on one curve, which indicates that the differences in the magnetoresistance are not solely caused by the scattering ($\tau$). So the LMR in ZrV$_6$Sn$_6$ is not a trivial result in the intermediate region. Instead, it is caused by other physical effects and requires further discussion.

    Parish and Littlewood \cite{PL} propose that the macroscopic inhomogeneity of the system \cite{PLexample2012,PLexample2020} will lead to a mixing of the longitudinal and transverse resistance . The linear Hall resistance of a single band in the transverse direction will result in the longitudinal LMR \cite{PLdiscuss1}. Nevertheless, in ZrV$_6$Sn$_6$, the Hall resistance is non-linear at low temperatures as shown in Fig.\textcolor{blue}{S4}. Moreover, as shown in Fig.\textcolor{blue}{3(d)}, there are obvious differences in the values of the transverse resistance and the longitudinal resistance, and longitudinal resistance exceeds the upper limit of Parish–Littlewood model ($\Delta\rho_{xx}< \Delta\rho_{xy}$) at low temperature \cite{QLMR2d2}. Furthermore, the LMR obtained from this model should be isotropic \cite{PLdiscuss2}, but the magnetoresistance behaviors in ZrV$_6$Sn$_6$ under different types vary significantly. Therefore, the LMR in ZrV$_6$Sn$_6$ is not caused by the Parish–Littlewood model.

    \begin{figure}
\includegraphics[width=0.9\linewidth]{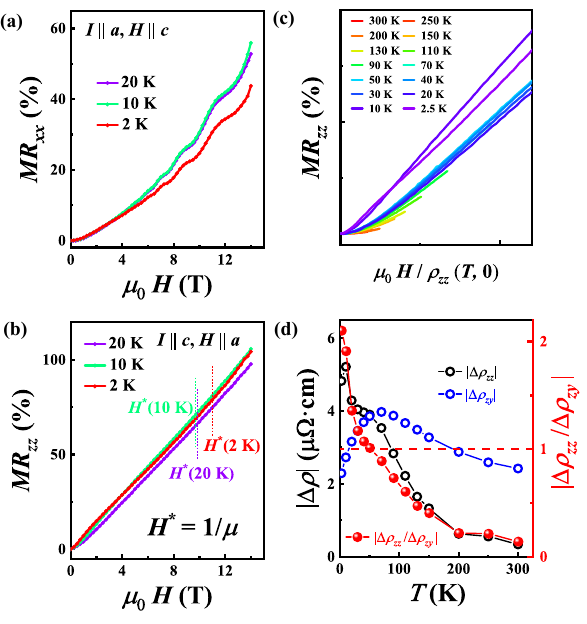}
\begin{center}
\caption{\label{1} The high-field magnetoresistance at low temperature in Configuration A (a) and Configuration B (b). (c): A plot of $MR_{zz}$ versus $\mu_{0}H/\rho_{zz}$($T$,0) at different temperatures. (d): The temperature-dependence of increments of longitudinal resistance (black circles) and the transverse resistance (blue circles) at 8.5 T, and their ratio (red dots).}
\end{center}
\end{figure}

    Abrikosov propose that in a system with linear energy-momentum dispersion, when electrons are filled into the first Landau level and reach the quantum limit, quantum LMR will occur \cite{QLMR}. Abrikosov model only discusses the case of 3D linear dispersion. The quantization can occur in the presence of a field in any direction shown as in Fig.\textcolor{blue}{4(a)} and Fig.\textcolor{blue}{4(c)}. The momentum direction $\gamma$ can be in any direction, $\alpha$ and $\beta$ correspond to it. We can observe the LMR in some 3D Dirac dispersion systems \cite{QLMR3d1,QLMR3d2}. However, it is not sufficient to explain why the LMR only appears in Configuration B of ZrV$_6$Sn$_6$.

    The LMR of ZrV$_6$Sn$_6$ obtained in our experiments exhibits significant anisotropy, and this anisotropy is difficult to explain using other existing models. Here, we point out that the 2D special case of quantum LMR can provide a plausible explanation for both the LMR in ZrV$_6$Sn$_6$ and its anisotropy. There are Dirac dispersion under the tight-binding model in Kagome lattice \cite{kagomeband} or topological nodal ring \cite{ZrV6Sn6} in ZrV$_6$Sn$_6$, which lead to energy bands with linear dispersion that depends only on $p_x$ and $p_y$ shown in Fig.\textcolor{blue}{4(b)}.
    Abrikosov considered the Hamiltonian of Dirac linear dispersion:

\begin{equation}
 \mathcal{H} = \int dV \ \psi^+ [v\boldsymbol{\sigma}(\mathbf{p}-\frac{e}{c} \mathbf{A})]\psi
\end{equation}

    Where $v$ is the Fermi velocity, $\boldsymbol{\sigma}$ is the Pauli operator and so \textbf{A} is the vector potential. In our system, we consider a 2D Hamiltonian around the linear dispersion, assuming that the $z$ is independent of $x$ and $y$. The system will undergo Landau quantization under Configuration A and Configuration B, respectively. (The detailed derivation is provided in the Supplementary Materials \cite{SM})

The energies in Configuration A:

\begin{align}
\varepsilon_{n}^+ &= v\sqrt{\frac{2eHn}{c}} \\
\varepsilon_{n}^- &= -v\sqrt{\frac{2eHn}{c}}
\end{align}

   The energies corresponding to any $n$ ($n$ = 0, 1, 2, $\ldots$) are independent of $p_x$ and $p_y$. They can be shown in Fig.\textcolor{blue}{4(d)}.

    The energies in Configuration B:
\begin{align}
\varepsilon_n^+ &= v\sqrt{p_x^2+\frac{2eHn}{c}} \\
\varepsilon_n^- &= -v\sqrt{p_x^2+\frac{2eHn}{c}}
\end{align}

    The energy in Configuration B is shown in Fig.\textcolor{blue}{4(c)},
    it is equivalent to the 3D model in Abrikosov's theory \cite{QLMR}. Therefore, his model can be used for further explanation. When in the first Landau level with $n = 0$, LMR in Configuration B of ZrV$_6$Sn$_6$ can be obtained:

\begin{align}
\rho_{zz}& = \frac{1}{2\pi}\left(\frac{e^{2}}{\varepsilon_{\infty}v}\right)^{2}\ln\varepsilon_{\infty}\frac{N_{i}}{ecn_{0}^{2}}H  \propto H
\end{align}

    The energy in Configuration A is at the first Landau level with $n$ = 0, the energy is constantly zero. Therefore, the change in energy in the $z$-direction cannot be regarded as a small quantity. For the transport problem, \(\varepsilon(z)\) must be taken into account, and the energy spectrum of a normal metal should still be satisfied in the $z$-direction. As a result, the quadratic MR of general metals is still maintained in Configuration A.
    
\begin{figure}
\includegraphics[width=1\linewidth]{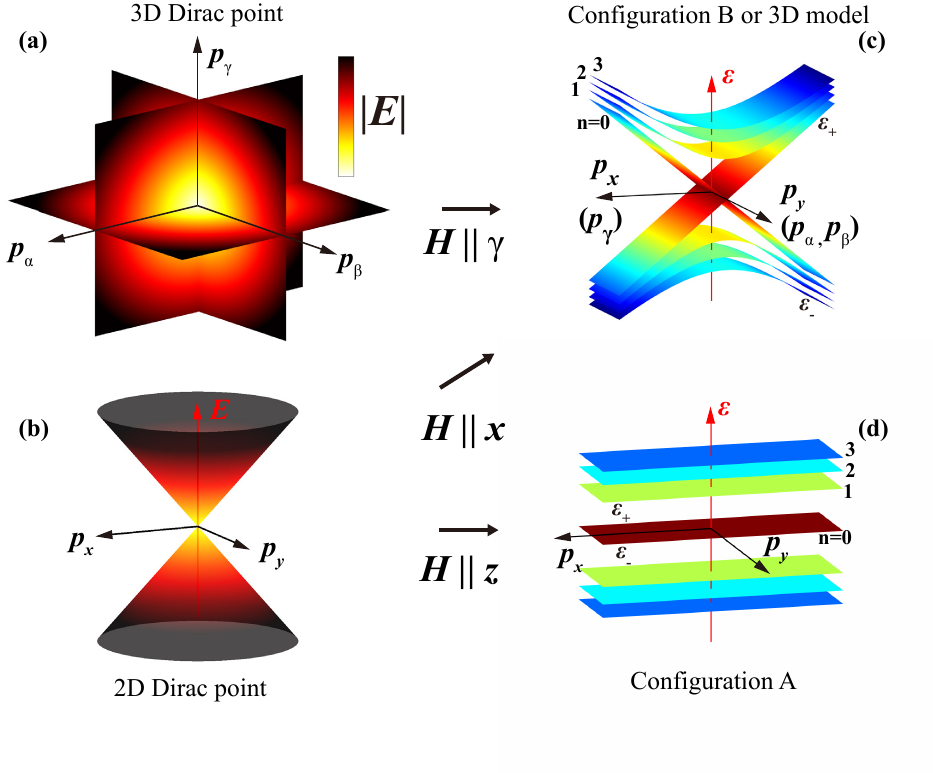}
\begin{center}
\caption{\label{1} (a), (b): Band structure of the 3D Dirac point and the 2D Dirac point, respectively. (c): The quantized energy-momentum dispersion of the 3D Dirac point when the field is along a certain direction $p_\gamma$. (d), (e): The quantized energy-momentum dispersion of the 2D Dirac point when the field is along $p_z$ for Configuration A and $p_x$ for Configuration B, respectively.}
\end{center}
\end{figure}	

    Furthermore, due to the rotational symmetry of the 2D linear dispersion, quantization of Configuration B can be formed when the field is in any direction within $ab$-plane. The LMR can be formed when the field is in any direction within the $ab$-plane as shown in Fig.\textcolor{blue}{2(f)} and Fig.\textcolor{blue}{2(g)}. However, when the field rotates within a plane perpendicular to the $ab$-plane, the LMR only appears when the field lies within the $ab$-plane, as depicted in the Fig.\textcolor{blue}{2(e)} and Fig.\textcolor{blue}{2(g)}. The field rotation experiments further demonstrate the correctness of our theory. The 2D model of quantum LMR can match the experimental results very well. And our model expands the Abrikosov model, making it more universal in explaining the LMR in linear dispersion systems. These works reveal the connection between the nontrivial electronic structures and the macroscopic quantum behavior in the Kagome system. Whether this 2D  quantum LMR can emerge in other 2D linear dispersion systems is worthy of further verification and study.

In summary, we systematically investigated the anisotropic transport behavior of the Kagome metal ZrV$_6$Sn$_6$. The LMR was discovered under specific directions of current and field. Furthermore, the LMR begins to manifest in the high field region below 90 K, subsequently extending towards lower field as temperature decreases further. Combining experiments with theories, we attribute the anisotropic LMR to the 2D scenario of the Abrikosov model. Our research provides an alternative explanation for the LMR and demonstrates that ZrV$_6$Sn$_6$ is an excellent platform for exploring the anomalous physical properties in Kagome system.

	This work was supported by the National Key Research and Development Program under Contracts of China (Grants No. 2021YFA1600201 and No. 2023YFA1607402), the National Natural Science Foundation of China (Grants No. 12274412, No. 12204487 and No. U2032215), the Open Research Fund of the Pulsed High Magnetic Field Facility (Grant No. WHMFC2024025), Huazhong University of Science and Technology, and the Systematic Fundamental Research Program Leveraging Major Scientific and Technological Infrastructure, Chinese Academy of Sciences, under Contract No. JZHKYPT-437 2021-08.

\bibliographystyle{apsrev4-2-title}
\bibliography{DYF}
\end{document}